\documentclass[aps,amssymb,prl, amsmath, twocolumn, floatfix]{revtex4}
\usepackage{graphics}
\usepackage{psfrag}
\usepackage{epsfig}
\usepackage{subfigure}
\usepackage{amssymb,amsfonts,amsmath}

\begin{document}



\title{Reliable protein folding on non-funneled energy landscapes: the free energy reaction path}
\author{Gregg Lois}
\author{Jerzy Blawzdziewicz}
\author{Corey S. O'Hern} 

 \affiliation{Department of Physics, Yale University, New Haven, CT 06520-8120;  Department of Mechanical Engineering, Yale University, New Haven, CT 06520-8286}


\begin{abstract} 
A theoretical framework is developed to study the dynamics of
protein folding.  The key insight is that the search for the native
protein conformation is influenced by the rate $r$ at which
external parameters, such as temperature, chemical denaturant or pH, are adjusted to induce folding.  A theory based on
this insight predicts that (1) proteins with non-funneled energy
landscapes can fold reliably to their native state, (2) reliable folding can occur as
an equilibrium or out-of-equilibrium process, and (3) reliable folding only
occurs when the rate $r$ is below a limiting value, which can be
calculated from measurements of the free energy.  We test these
predictions against numerical simulations of model proteins
with a single energy scale.
\end{abstract}
\maketitle

Under appropriate conditions, proteins spontaneously fold from an extended one-dimensional chain of amino acids to a unique three-dimensional native conformation.  
How this occurs on timescales accessible to experiment---and relevant to biological function---is a question that has intrigued scientists for the past forty years.  Levinthal~\cite{levinthal} was the first to recognize the importance of timescales and point out that, assuming a random search of conformation space, proteins would not fold in a person's lifetime.  This argument has come to be known as Levinthal's Paradox since proteins must fold for human life to exist in the first place.

Of course conformation space is not sampled randomly and Levinthal's paradox has been resolved by applying statistical mechanics to the protein folding problem~\cite{zwanzig,dillnature,plotkinfunnel}.  Each protein conformation has a free energy that determines its probability to be sampled at temperature $T$.  While the free energy $F$ generally comprises a sum of many enthalpic and entropic terms, it is convenient to express it as $F=E-TS_\mathrm{conf}$, where $S_\mathrm{conf}$ is the conformational entropy of only the protein degrees of freedom and $E$ is the ``internal energy'' that includes all other contributions to the free energy (from both protein and solvent).  The functional dependence of $E$ on all protein degrees of freedom is called the energy landscape~\cite{landscapebook,dillenergysim}, which in general contains many minima.  At $T=0$ only the energy landscape is relevant and the protein resides in a local (or global) minimum, corresponding to a compact conformation.  As $T$ increases the conformational entropy smooths out the minima in the energy landscape and the protein adopts more extended states with larger $S_\mathrm{conf}$.  In the ``new view'' of protein folding~\cite{dillnature, baldwinnewview} statistical fluctuations on an energy landscape give rise to an ensemble of folding pathways. 

Often associated with the new view is the hypothesis that energy landscapes have the shape of a multi-dimensional funnel~\cite{plotkinfunnel,otherfunnel}.  Proponents argue that in order to fold reliably (transition to the native state with probability one) the energy landscape must contain a single low-lying minimum to which all conformations are channeled.  If multiple funnels exist, separated by large enough energy barriers, then at low temperature or denaturant concentration a protein can become trapped in a local minimum of energy that does not correspond to its native conformation.  While the existence of a single funnel is a sufficient condition for reliable protein folding, the number of proteins with a single funnel is expected to be small and the observation of kinetic traps~\cite{kinetictraps} and glassy behavior~\cite{glassiness} in biologically relevant proteins indicates that not all proteins fold on smooth funneled landscapes.

Here we address the open question:  is a funneled energy landscape necessary for reliable folding?  By formulating a statistical theory that includes the dynamics of folding, we find that a funneled landscape is not necessary for reliable folding.  The important insight is that the rate $r$ at which temperature or chemical denaturant concentration is decreased to induce folding affects the final conformation of the protein.  For sufficiently small $r$ the protein always folds to its native conformation, whereas for larger $r$ it can become trapped in a metastable state.  This leads to new predictions that can be tested in experiments and simulations.  First, proteins with non-funneled energy landscapes can fold reliably to their native state if the rate $r$ is below a limiting value.  Second, reliable folding can occur as an equilibrium-quasistatic or non-equilibrium process.  Third, in a non-equilibrium folding process, a protein can reliably fold to a local (instead of global) minimum of the energy landscape.  We conduct off-lattice simulations of model proteins with non-funneled energy landscapes and verify these predictions.  

\section{Results}
We consider proteins with general energy landscapes---not necessarily funneled---and derive the conditions under which folding occurs reliably.  Generally, energy landscapes contain multiple minima, possibly separated by large energy barriers.  Thus folding is not necessarily an equilibrium process and misfolds can occur.  Below we consider the dynamics of the folding process and its effect on reliable folding.

\subsection{A kinetic mechanism for folding}
Multiple minima in the energy landscape lead to multiple minima in the free energy.  In this case we argue that there is a basic kinetic mechanism that determines whether folding is reliable.  We illustrate this kinetic mechanism by considering a transition from state ${\bf A}$ to state ${\bf B}$ on a non-funneled energy landscape.  Although we will assume that the transition is driven by a reduction of temperature, the same arguments can be applied when a change of denaturant concentration or another parameter induces folding.

\begin{figure}
\begin{center}
\begin{tabular}{cc}
\scalebox{0.38}{\includegraphics{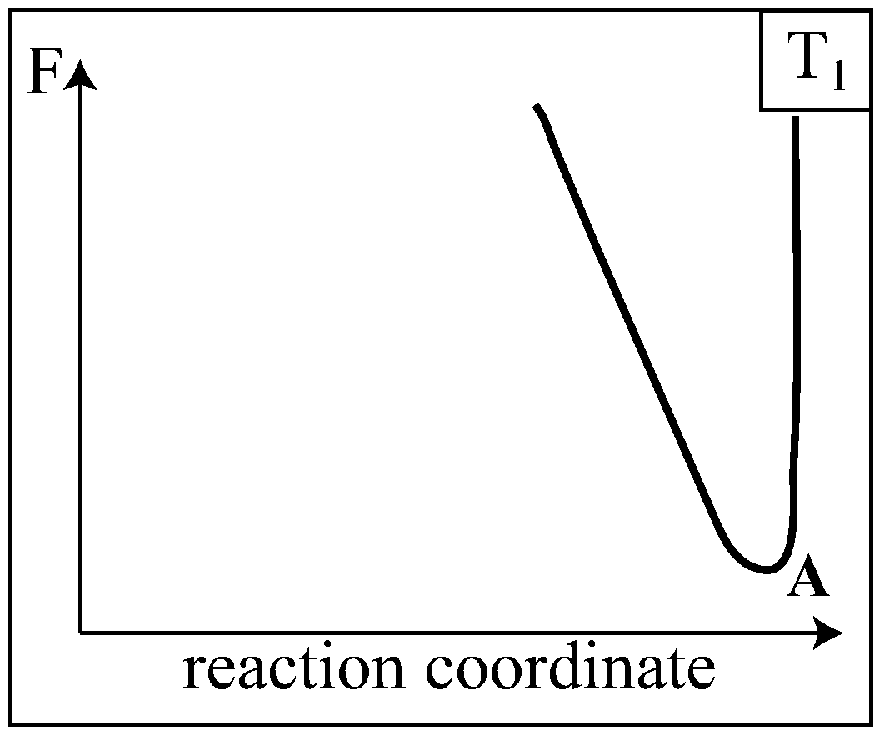}}
&
\scalebox{0.38}{\includegraphics{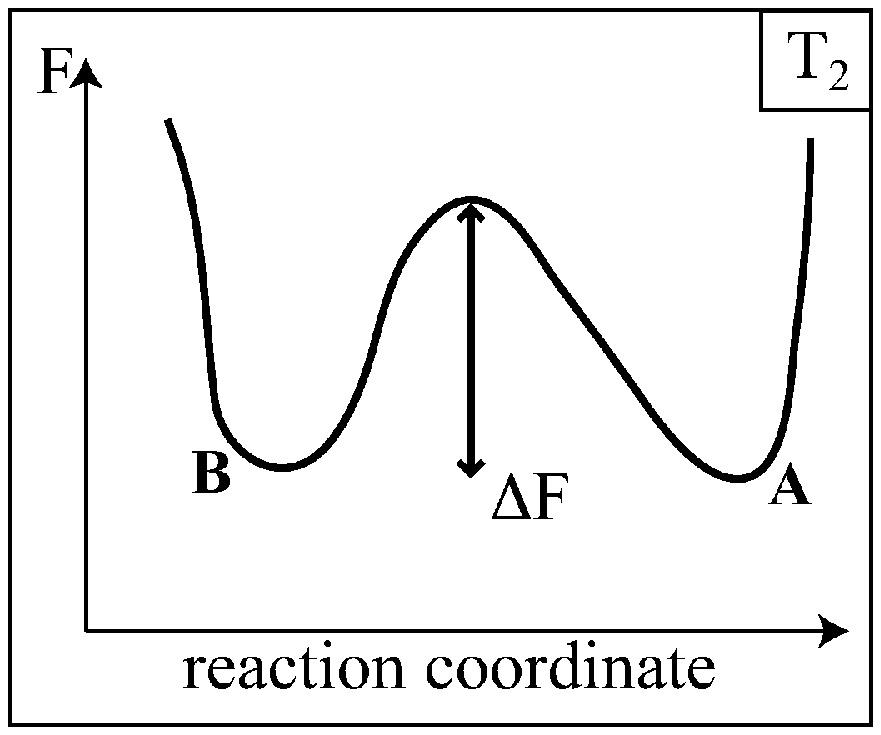}}
\\
\scalebox{0.38}{\includegraphics{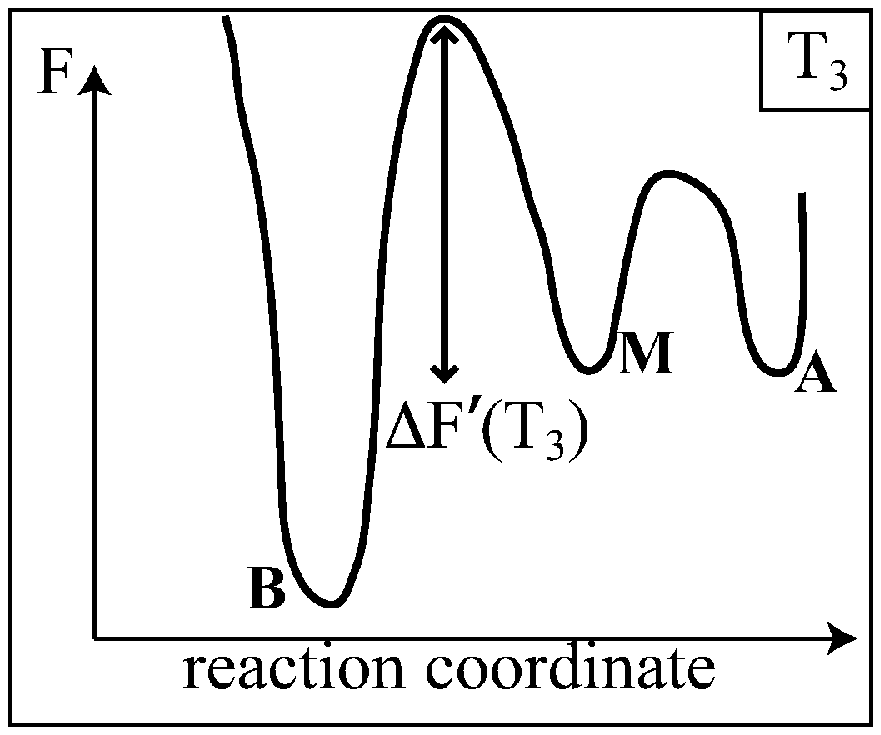}}
&
\scalebox{0.38}{\includegraphics{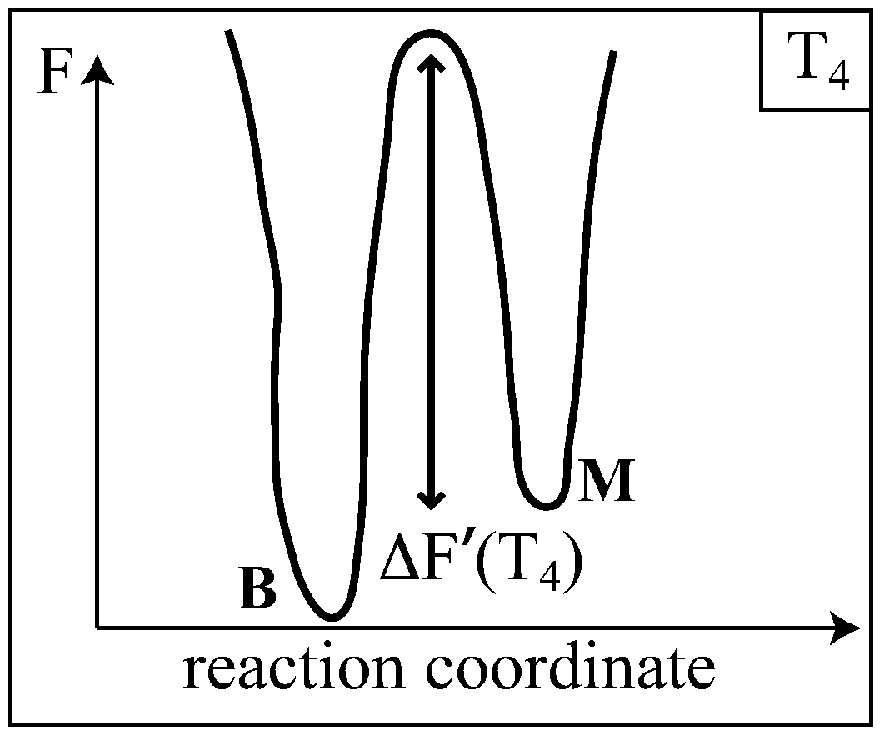}}
\end{tabular}
\end{center}
\caption{\label{schematicfreeenergy}  Schematic plots of the free energy versus an arbitrary reaction coordinate at four temperatures where $T_1 > T_2 > T_3 > T_4$.  At $T_1$ only the state ${\bf A}$ is accessible.  At $T_2$, transitions to state ${\bf B}$ occur with activation barrier $\Delta F$.  $T_3$ is defined as the largest temperature at which a new state ${\bf M}$ exists with free energy equal to that of state ${\bf A}$.  If the protein has not transitioned to state ${\bf B}$ by $T_3$, misfolds can occur.  At $T_4$ the free energy barrier $\Delta F'(T)$ separating ${\bf M}$ and ${\bf B}$ becomes larger than it was at $T_3$.
}
\end{figure}

In Fig.~\ref{schematicfreeenergy} schematic illustrations of the free energy are plotted at four temperatures $T_1>T_2>T_3>T_4$.  We will assume that a transition from ${\bf A} \rightarrow {\bf B}$ is induced by decreasing the temperature at a constant rate $r$ such that $T(t)= T_1(1-rt)$ as a function of time $t$.  Initially at $T_1$ the protein resides in state ${\bf A}$.  As temperature is reduced to $T_2$ an equilibrium transition to state ${\bf B}$ can occur with folding time proportional to $\exp(\Delta F/T_2)/r^*$, where $r^*$ is the rate at which conformations are explored.  At $T_3$ a third state ${\bf M}$ has free energy equal to that of ${\bf A}$.  As temperature is further reduced to $T_4$, the minimum corresponding to state ${\bf A}$ no longer exists and the activation barrier $\Delta F'$ grows.  

Dynamics are important in determining transitions between states ${\bf A}$ and ${\bf B}$.  If the time that it takes for the temperature to decrease from $T_2$ to $T_3$ is less than the folding time, the protein can fall into the metastable state ${\bf M}$.  This sets a bound on $r$:  if             
\begin{equation}
r>r^f \equiv \frac{(T_2-T_3) r^*}{T_1} \exp(\frac{-\Delta F}{T_2})
\label{tempfall}
\end{equation}
then the protein is likely to populate the state ${\bf M}$.  Note that we use units where Boltzmann's constant $k_B=1$.

For a misfold to occur, the escape probability from the metastable state must be sufficiently small.  If the protein populates state ${\bf M}$ at time $t_3$, the probability that it has escaped at time $t$ is given by 
\begin{eqnarray}
P(t-t_3) &=& \exp\left(-\int_{t_3}^t dt \,r^* \exp(-\Delta F'(T)/T) \right) \nonumber \\
&\equiv& \exp\left(-g(t-t_3)\right). 
\label{residencetime}
\end{eqnarray}
For a maximum waiting time $\tau$ the protein always escapes the metastable state for $g(\tau) \gg 1$ and rarely escapes for $g(\tau) \ll 1$.  The crossover between frequently escaping from and being trapped in state ${\bf M}$ occurs when $g(\tau) \approx 1$.     
Using $T(t)=T_1(1-rt)$ we find that when the rate
\begin{equation} 
r > r^s \equiv \int_0^{T_3} r^* \, \exp(\frac{-\Delta F'(T)}{T}) \, \frac{dT}{T_1},
\label{tempstuck}
\end{equation} 
the probability to become trapped in the metastable state ${\bf M}$ is significant and misfolds occur~\footnote{Here we use a waiting time $\tau$ that satisfies $T(\tau)=0$.  Note that the limiting rates $r^f$ and $r^s$ can also be determined for any functional form $T(t)$ (which we assumed to be linear) and maximum waiting time $\tau$.}.

From these basic considerations it is apparent that protein folding transitions on non-funneled energy landscapes are influenced by multiple minima in the free energy and the rate $r$ at which external parameters are varied to induce folding.  To determine whether reliable folding occurs we must address two important questions:  (i) can the protein conformation reside in a metastable local minimum? and (ii) is it likely that the protein conformation becomes trapped in that local minimum?  The answers to these questions define the limiting rates $r^f$ and $r^s$.  The transition ${\bf A} \rightarrow {\bf B}$ occurs reliably if $r$ obeys one of the inequalities, $r<r^f$ or $r<r^s$.  In the case that $r<r^s$ the protein is given sufficient time to sample all states and the transition ${\bf A} \rightarrow {\bf B}$ occurs reliably as an equilibrium process.  If $r^s<r<r^f$ the protein conformation becomes trapped in the state ${\bf B}$ without fully exploring phase space and the transition occurs reliably, but out of equilibrium.  If $r>r^f$ and $r>r^s$ then the protein does not transition between ${\bf A}$ and ${\bf B}$ reliably.

\subsection{The Free Energy Reaction Path}
In the previous section we identified a kinetic mechanism that influences transitions on non-funneled landscapes.  In this section we use this mechanism to formulate a general framework for understanding folding.  We begin by partitioning the energy landscape into basins associated with particular protein topologies, proceed to define the free energy reaction path that describes how the protein transitions from one topology to another, and then use the kinetic mechanism described above to determine whether folding is reliable.

As a way to understand complex folding dynamics, 
the energy landscape of an arbitrary protein can be partitioned into basins surrounding each local minimum, analogous to the inherent structure formalism for liquids and glasses~\cite{stillinger}.  In particular, the infinite number of protein conformations can be uniquely associated with a finite number of topologies, defined as protein conformations that correspond to local minima of the internal energy.  We denote a topology as $\mathbf{t}^n$, where $n$ is an index that contains sufficient information to fully describe the conformation ({\em e.g.} number, type and arrangement of bonds).  The set of conformations $\mathcal{B}(\mathbf{t}^n)$ associated with each topology $\mathbf{t}^n$ is the basin of attraction for that topology.  The basin of attraction is defined such that all conformations that belong to $\mathcal{B}(\mathbf{t}^n)$ relax to the topology $\mathbf{t}^n$ when thermal fluctuations of the protein are suppressed.  Thus the infinite number of possible protein conformations is represented by a finite number of topologies and a free energy $F(\mathbf{t}^n)$ can be defined for the set of protein conformations $\mathcal{B}(\mathbf{t}^n)$.  Formally the partition function $Z(\mathbf{t}^n)$ for conformations constrained to lie in $\mathcal{B}(\mathbf{t}^n)$ is given by
\begin{equation}
Z(\mathbf{t}^n) = \int_{\mathcal{B}(\mathbf{t}^n)} \exp(-E/T)\, d\Gamma,
\end{equation}  
where integration is over all coordinates $\Gamma$ in the basin $\mathcal{B}(\mathbf{t}^n)$ and $E$ is the internal energy as a function of $\Gamma$.  The free energy for a protein constrained to $\mathcal{B}(\mathbf{t}^n)$ can then be written in terms of the topology $\mathbf{t}^n$ as 
\begin{equation}
F(\mathbf{t}^n,T) = E(\mathbf{t}^n,T) - T S_\mathrm{conf}(\mathbf{t}^n,T),
\end{equation}
where $E(\mathbf{t}^n,T)$ is the internal energy of topology $\mathbf{t}^n$ and $S_\mathrm{conf}(\mathbf{t}^n,T)$ is its associated entropy~\cite{stillinger}, given by
\begin{equation}
S_\mathrm{conf}(\mathbf{t}^n,T) = \log \int_{\mathcal{B}(\mathbf{t}^n)} \exp\left(-\big[E-E(\mathbf{t}^n,T)\big]/T \right)\, d \Gamma .
\end{equation}
The random coil state $\mathbf{t}^0$ with zero internal energy has the largest entropy and is therefore the global minimum of free energy at sufficiently large temperature.  

Given a protein with an energy landscape that has been partitioned into basins of attraction, we define the free energy reaction path as the ordered sequence of topologies that the protein adopts as temperature is reduced in the equilibrium limit.  That is, if the rate $r$ is sufficiently small, the protein will come to equilibrium at all temperatures and proceed through the basins of attraction for a reproducible set of topologies $\mathbf{t}^0 \rightarrow \mathbf{t}^{n_1} \rightarrow \mathbf{t}^{n_2} \rightarrow \cdots \rightarrow \mathbf{t}^{n_{N}}$.  Each transition occurs at the temperature where the free energy of two topologies is equal, {\em e.g.} the transition $\mathbf{t}^0 \rightarrow \mathbf{t}^{n_1}$ occurs at the temperature $T^*$ where $F(\mathbf{t}^0,T^*) = F(\mathbf{t}^{n_1},T^*)$.  In this way, for any energy landscape, the free energy reaction path encodes the path taken through conformation space when folding occurs as an equilibrium-quasistatic process.

To determine whether folding is reliable, we apply the analysis introduced in the previous section to each transition in the free energy reaction path.  If we label the transitions by $i=1,2,\ldots,N$ then limiting rates $r^f_i$ and $r^s_i$ can be determined for each transition by measuring properties of the free energy.  There are then three distinct folding scenarios:  (1) if $r<r^s_i$ for all $i$ then the protein does not become trapped in metastable conformations and folding occurs reliably in equilibrium; (2) if $r^s_i < r < r^f_i$ for a single transition $i$ then the protein falls out of equilibrium at transition $i$, but reliably folds to the topology $\mathbf{t}^{n_i}$ (since the condition $r<r^f_i$ guarantees that the protein does not fall into a different metastable state).  Note that if there exist multiple transitions with $r^s_i<r<r^f_i$ then the protein will reliably fold to the topology with the smallest value of $n_i$ for which this condition holds. Finally, (3) if $r>r^s_i$ and $r>r^f_i$ for any $i$, and condition (2) does not hold for a smaller value of $i$, then the protein will not fold reliably.

From our analysis we deduce that there are two types of reliable folding, equilibrium and non-equilibrium.  While reliable equilibrium folding brings the protein to the global minimum of free energy, reliable non-equilibrium folding can target local minima.  The free energy reaction path provides a useful framework to classify the relevant transitions since, depending on the rate $r$, a protein will either (1) pass through all topologies on the free energy reaction path and arrive at the topology with the smallest free energy, (2) target an intermediate topology along the free energy reaction path and reliably fold to a local minimum of free energy, or (3) misfold and deviate from the free energy reaction path.  

\subsection{Simulations of a model protein}
\begin{figure}
\begin{center}
\scalebox{0.60}{\includegraphics{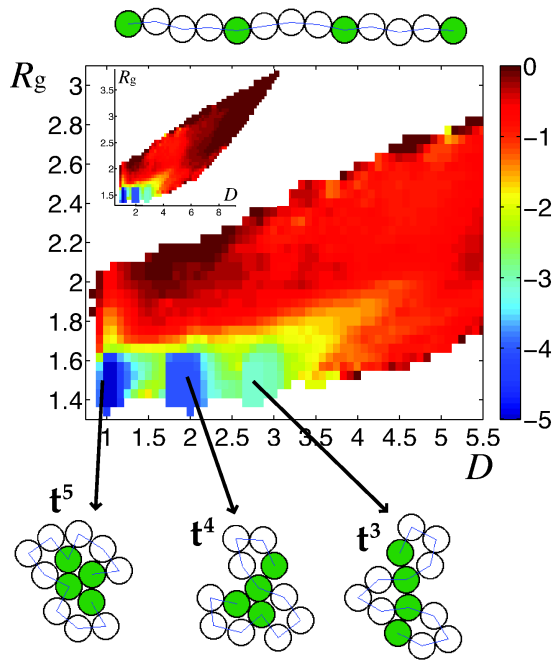}}
\end{center}
\caption{\label{energylandscape} Contour plot of the energy landscape and pictures of the relevant topologies for a model protein.  The fully extended conformation is shown at the top of the figure.  The inset displays the full energy landscape and the main figure contains a magnified view of the compact states.  The landscape is plotted as a function of the radius of gyration $R_g$ and end-to-end distance $D$, each normalized by the monomer diameter.  The colorbar gives the total internal energy of the protein divided by the attraction strength $|E_c|$.  There are three distinct energy minima separated by barriers and the associated topologies are pictured.
White regions correspond to protein conformations that are never sampled in the simulations.
}
\end{figure}

To test the predictions of the previous section we perform off-lattice Brownian dynamics simulations of a model protein with a single attractive energy scale.  We model the protein as a polymer chain containing both attractive (green) and non-attractive (white) spherical monomers of size $\sigma$.  Interactions between non-adjacent green monomers are attractive with energy depth $E_c<0$, while interactions between non-adjacent pairs of green-white or white-white monomers are purely repulsive.  This model is a variant of the ``HP'' model~\cite{hpmodels}.  Thermal fluctuations of the protein at temperature $T$ are included using Brownian dynamics simulations with solvent viscosity $\eta$.  We observe that as the parameter $c=|E_c|/T$ increases from zero the polymer chain transitions from a random coil to a folded conformation.  To test the predictions of the theory we simulate a specific sequence of green and white monomers, pictured in Fig.~\ref{energylandscape}.  In this article we present results for two dimensions in order to simplify identification of the multiple topologies that the polymer chain adopts.  We have also conducted simulations in three dimensions and these results are included in the supporting information.

In Fig.~\ref{energylandscape} we plot the energy landscape of the polymer chain as a function of two reaction coordinates: the radius of gyration $R_g$ and the end-to-end distance $D$, each normalized by the monomer diameter $\sigma$.  In terms of these two reaction coordinates, three energy minima exist and are separated by energy barriers.  The minima correspond to three distinct topologies that are pictured in Fig.~\ref{energylandscape}.  We find a total of four relevant topologies for this simple system, containing either zero $\mathbf{t}^0$, three $\mathbf{t}^3$, four $\mathbf{t}^4$, or five $\mathbf{t}^5$ bonds between attractive green monomers.  Energy barriers exist between $\mathbf{t}^3$, $\mathbf{t}^4$ and $\mathbf{t}^5$ because, in order to transition between the topologies, it is necessary to first break a bond and then rearrange the chain conformation.  Note that four green particles is the minimum number needed to ensure multiple energy minima in two dimensions, while seven are required in three dimensions.  Including additional green particles introduces additional minima and more complicated energy landscapes---we treat only the simplest case here. 

\begin{figure*}
\begin{center}
\mbox{
\scalebox{1}{\includegraphics{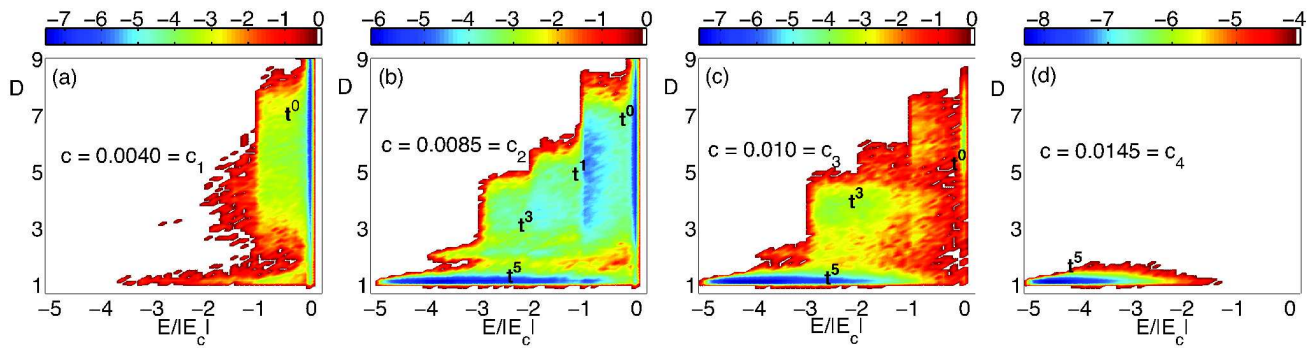}}
}
\end{center}
\caption{\label{figfreeenergya}  Contour plots of the free energy $F/T$ normalized by temperature as a function of $E/|E_c|$ (horizontal axis) and end-to-end distance $D$ (vertical axis) for a sequence of $c$-values.  The free energy is calculated from the probability for the protein to be in a conformation with given $E/|E_c|$ and $D$.  White regions correspond to protein conformations that are never sampled in the simulations.     
}
\end{figure*}

Given the non-funneled energy landscape of the simulated protein we now determine the associated free energy reaction path.  Measurements of free energy $F/T$, normalized by temperature, as a function of $E/|E_c|$ and end-to-end distance $D$ are shown in Fig.~\ref{figfreeenergya} for a sequence of $c$-values that corresponds to the sequence of schematic plots in Fig.~\ref{schematicfreeenergy}.  In Fig.~\ref{figfreeenergya}(a) we plot $F/T$ for a small value of $c=0.0040$ and observe that the random coil state $\mathbf{t}^0$ is the only free energy minimum.  In Fig.~\ref{figfreeenergya}(b) $c$ is increased to $c_2=0.0085$ and there are multiple local minima in the free energy, including the topologies $\mathbf{t}^0$, $\mathbf{t}^1$, $\mathbf{t}^3$, and $\mathbf{t}^5$.  The free energies of $\mathbf{t}^0$ and $\mathbf{t}^5$ are equal in Fig.~\ref{figfreeenergya}(b).  At a slightly higher value $c=c_3=0.0100$, Fig.~\ref{figfreeenergya}(c) exhibits three minima and the free energy of $\mathbf{t}^0$ and $\mathbf{t}^3$ are equal.  Finally at $c=0.0145$, the free energy plotted in Fig.~\ref{figfreeenergya}(d) exhibits a deep minimum at topology $\mathbf{t}^5$.  

From the plots in Fig.~\ref{figfreeenergya} we conclude that the first and only transition in the free energy reaction path is $\mathbf{t}^0 \rightarrow \mathbf{t}^5$ where the protein folds to its native conformation.  Although other local minima exist in the free energy and misfolds are possible for $c>c_3$, $F(\mathbf{t}^5)$ is the global minimum of free energy for $c>c_2$.  This simple polymer chain does not exhibit any intermediate states on the free energy reaction path, which prevents us from testing whether proteins can fold reliably to metastable minima.  However we will test all other predictions of the theory.  In the {\em Materials and Methods} section we calculate the limiting rates $r^f \eta \sigma^2/T= 1.8 \times 10^{-7}$ and $r^s \eta \sigma^2/T= 3.0 \times 10^{-8}$ for the single transition on the free energy reaction path, where $\eta \sigma^2/T$ is the simulation time-unit.

\begin{figure}[!h]
\begin{center}
\begin{tabular}{c}
{\bf (a)} \\
\mbox{
\psfrag{yl}{\Huge{$E/|E_c|$}}
\psfrag{xl}{\Huge{$c$}}
\scalebox{0.30}{\includegraphics{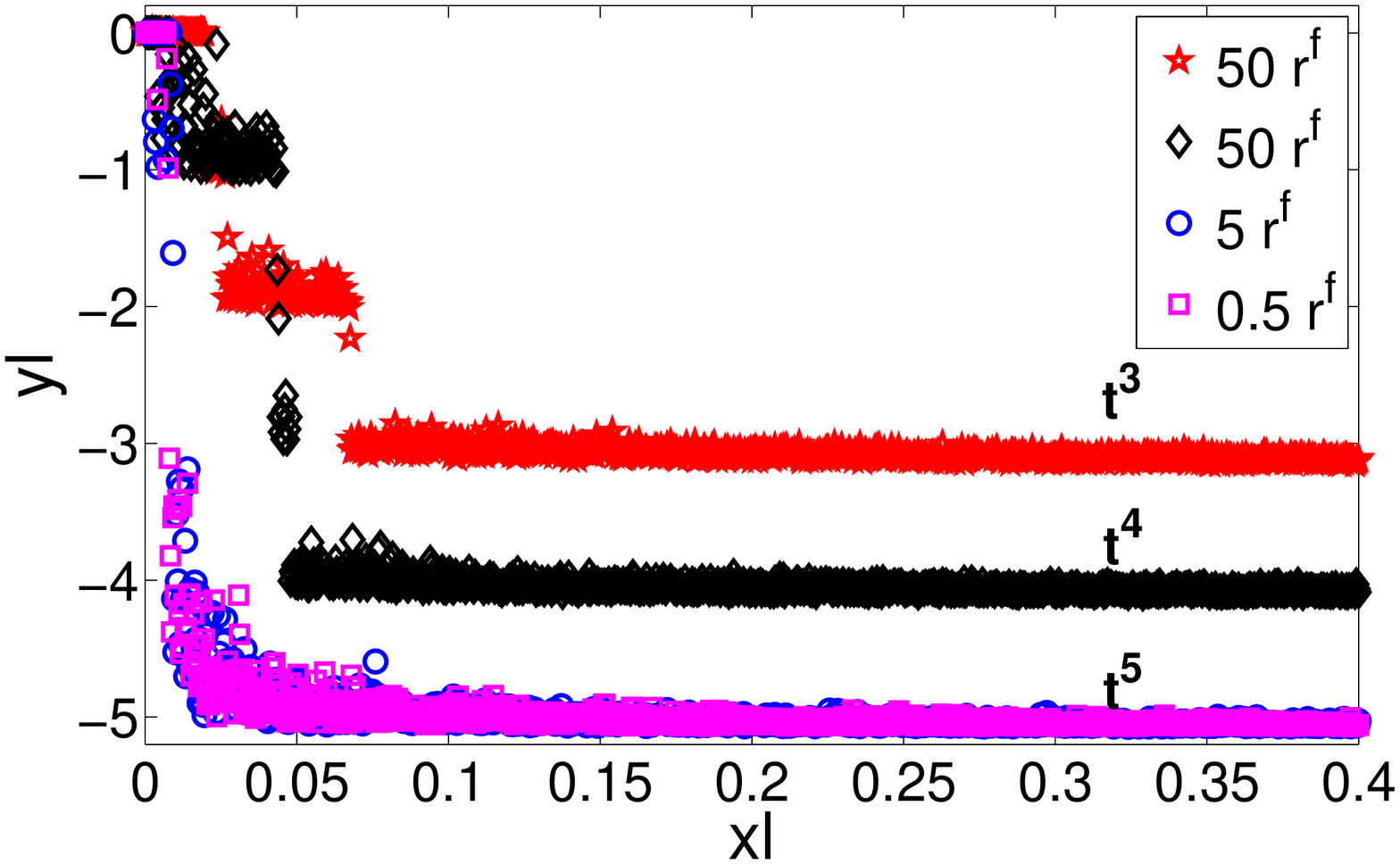}}
}
\\
{\bf (b)} \\
\mbox{
\psfrag{xl}{\Huge{$\log_{10}{(r\eta \sigma^2/T)}$}}
\psfrag{yl}{\Huge{$P_c$}}
\scalebox{0.30}{\includegraphics{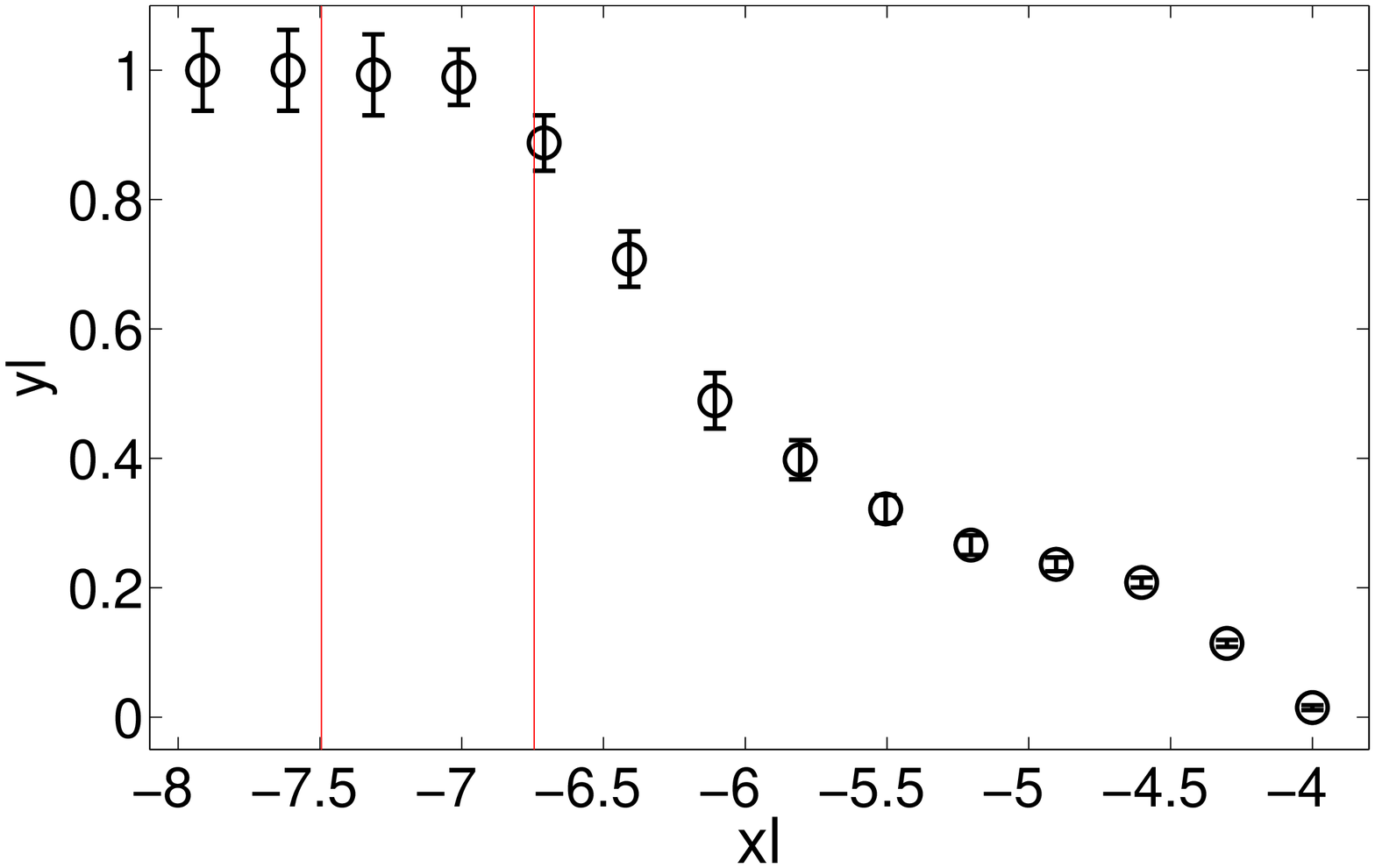}}
}
\\
{\bf (c)} \\
\mbox{
\psfrag{yl}{\Huge{$\delta E^2$}}
\psfrag{iyl}{\Huge{$\delta E^2$}}
\psfrag{xl}{\Huge{$c$}}
\psfrag{ixl}{\Huge{$\log_{10}(r/r^s)$}}
\scalebox{0.30}{\includegraphics{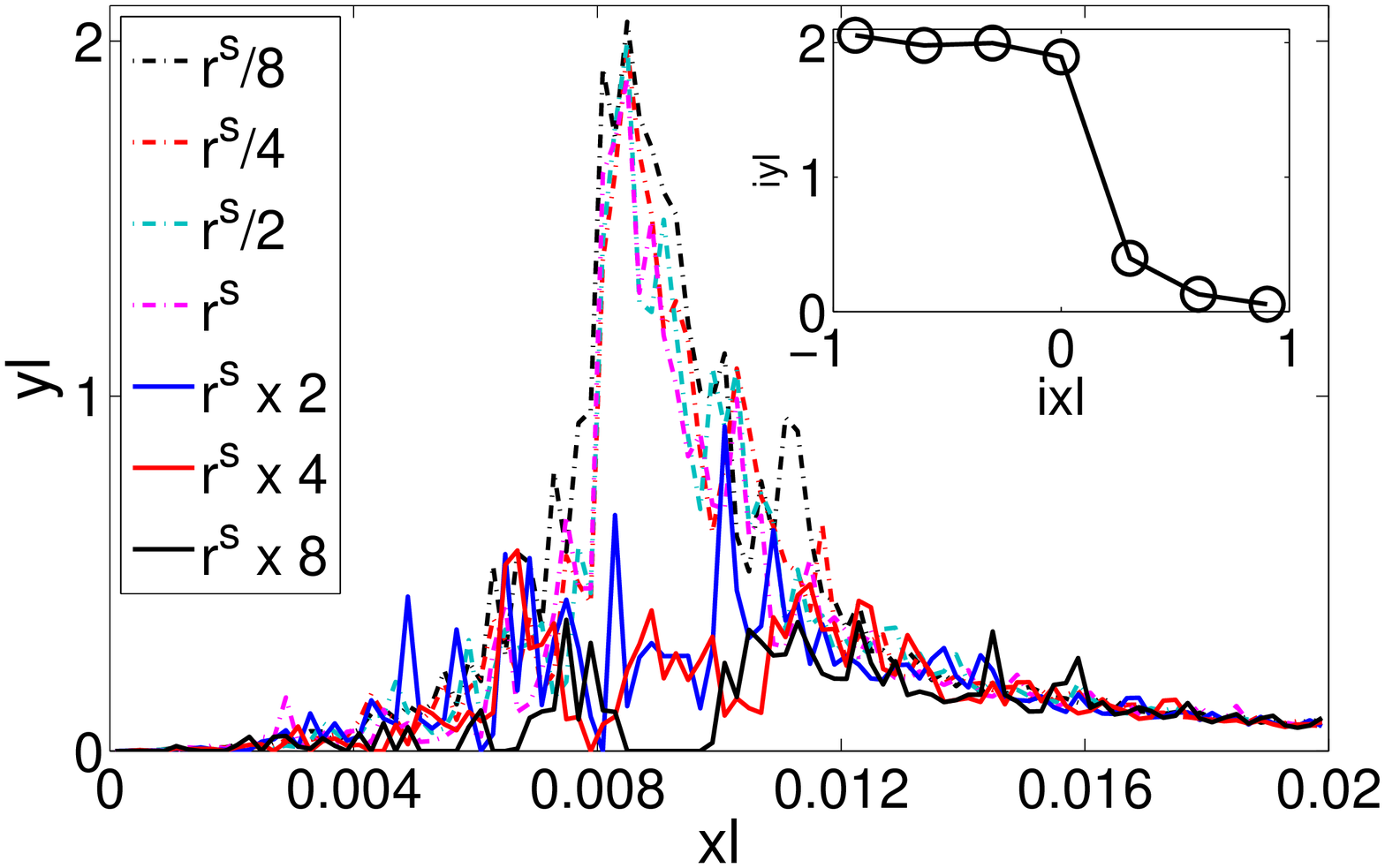}}
}
\end{tabular}
\end{center}
\caption{\label{figfolding}  {\bf (a)} Folding trajectories from simulations with identical initial conditions at three different rates.  The normalized energy $E/|E_c|$ is plotted as a function of c and the final state is labeled by its topology.  Slow rates lead to the native state $\mathbf{t}^5$ whereas fast rates lead to unreliable folding.  {\bf (b)} The probability of folding to the native state $P_c$ as a function of rate $r$.  Error bars are from sampling statistics.  For $r \eta \sigma^2 / T \lesssim 10^{-7}$ the protein folds reliably to the topology $\mathbf{t}^5$.  Vertical lines indicate the values of $r^f$ and $r^s$ calculated in the text. {\bf (c)} Energy fluctuations $ \delta E^2 = \big( \langle E^2 \rangle - \langle E \rangle^2\big)/E_c^2$ as a function of $c$ for folding simulations at different rates $r$.  For $r \leq r^s$ (dashed lines) the fluctuation curves appear to collapse and reliable folding occurs in equilibrium.  For $r^s<r<r^f$ (full lines) fluctuations depend on $r$ and reliable folding occurs out of equilibrium.  Inset:  Energy fluctuations at the equilibrium transition point $c=c_2=0.0085$ as a function of $r/r^s$.
}
\end{figure}

Now that we have determined the free energy reaction path and calculated the limiting rates, we conduct dynamic simulations of folding.
To induce folding in the polymer chain $c$ is increased linearly in time at rate $r$ ($c=rt$), starting from the topology $\mathbf{t}^0$ at $c=0$.  In Fig.~\ref{figfolding}(a) the energy of the polymer chain is plotted as a function of $c$ for three different values of $r$, with the final state labeled by its topology.  From this figure we clearly see that small $r$ targets the native state $\mathbf{t}^5$ whereas larger $r$ leads to misfolding.  In Fig.~\ref{figfolding}(b) we plot the probability to fold to the native state $\mathbf{t}^5$ as a function of $r \eta \sigma^2/T$, averaged over many folding trajectories studied for each $r$.  The protein folds reliably for small rates.  

The modern theory of protein folding requires funneled energy landscapes for reliable folding~\cite{plotkinfunnel,otherfunnel}.  The simple protein model we consider here provides a contradiction to this viewpoint since it does not possess a funneled landscape but nevertheless folds reliably at small $r$.  The free energy reaction path theory predicts that reliable folding can occur on non-funneled landscapes and provides a means to quantitatively determine the limiting rate below which folding is reliable. 
Given the values of $r^f$ and $r^s$ quoted above, the free energy reaction path theory predicts reliable folding for $r \eta \sigma^2/T < 1.8 \times 10^{-7}$.  In Fig.~\ref{figfolding}(b) we have measured that reliable folding occurs for normalized rates less than $\approx 10^{-7}$.   The theory therefore makes a correct quantitative prediction of the simulation results.  Additionally, the values of $r^f$ and $r^s$ indicate that there is a range of rates $r^s < r < r^f$ where reliable folding to $\mathbf{t}^5$ occurs out of equilibrium.  We test this prediction by measuring energy fluctuations for rates at which folding is reliable, as plotted in Fig.~\ref{figfolding}(c).  For $r \leq r^s$ fluctuations are large at the transition point $c=0.0085$ since the protein is sampling both folded and unfolded conformations as it remains in equilibrium.  For $r > r^s$ fluctuations remain small near the transition point since the protein becomes trapped in the folded state and reliable folding is a non-equilibrium process.

\section{Discussion}
Levinthal was the first to realize that the exponential number of collapsed conformations preclude a protein from finding its native state via random sampling.  The experimental observation that proteins fold reliably to a reproducible native state therefore requires an explanation.  The modern view is that protein sequences have evolved to favor energy landscapes with a single funnel and can therefore fold reliably.  We have demonstrated that proteins with non-funneled energy landscapes can also fold reliably, as long as the external parameters that induce folding are adjusted slowly enough.  

We have identified two reliable folding processes on non-funneled landscapes: equilibrium and non-equilibrium.  Even though it is possible that in experimental and biological settings the rate at which external parameters are varied to induce folding is too large to access the equilibrium limit, reliable folding can occur out of equilibrium.  If this is the case, the native state should be regarded as a reliably targeted local minimum on the free energy reaction path that remains metastable over timescales sufficient for biological function. 

The importance of the free energy reaction path and the necessity of using small rates to vary external parameters presents challenges for protein folding simulations.  Reliable protein folding is especially difficult to study in all-atom simulations where, due to the long time scales and large number of atoms, extremely rapid rates are used to induce folding~\cite{simfastrate}.  From our results, reliable folding on non-funneled landscapes depends on rate, thus simulation studies that argue that funneled energy landscapes are necessary for reliable folding~\cite{funnelsims} must be carefully interpreted if only large rates are considered.   

Our predictions can be tested in experiments by studying folding over a range of rates, using methods such as ultrafast mixing or laser pulsing~\cite{fastfoldingreview}.  Some progress has been made in this direction~\cite{fastpapers} and the observation of ``strange kinetics''~\cite{strangekinetics} after rapid temperature jumps is consistent with our predictions.  In three dimensions the limiting rates are proportional to $r^* \propto T/\eta R_H^3$, where $R_H$ is the hydrodynamic radius.  This implies that investigating folding in a variety of solvents with different viscosities $\eta$ can greatly increase the range of experimentally accessible rates.  Moreover, due to the inverse dependence on $T$, folding by changing temperature will give different limiting rates than folding by reducing denaturant concentration.  

Finally, it is intriguing to speculate about folding {\em in vivo}.  Given that the folded state of a protein is dependent on rate at which external parameters are varied to induce folding, and that local minima in free energy can be targeted by adjusting this rate, it is possible that protein sequence has evolved along with the biological environment in which it folds.  Since the folding process is determined by protein sequence and rate, both are likely used in nature to ensure robust folding.

\section{Materials and Methods}
\subsection{Simulation protocol}
Simulations are performed on polymer chains of spherical monomers, each with diameter $\sigma$.  We include two types of monomers---attractive (green) and non-attractive (white).  Interactions depend on the separation $r_{ij}$ between monomers $i$ and $j$, and it is convenient to define the normalized distance $\bar{r}_{ij} \equiv r_{ij}/\sigma$.  Interactions between adjacent monomers are chosen to prevent the polymer chain from breaking, while interactions between non-adjacent monomers are either purely repulsive (for green-white or white-white interactions) or attractive (for green-green interactions).  More specifically, monomers that are adjacent on the polymer chain experience a piecewise continuous potential $\Phi_{cc}(\bar{r})$ that is comprised of a purely repulsive Lennard-Jones (RLJ) potential~\cite{browniandynamicssim} for separations $\bar{r}_{ij} \leq 1$ and a FENE potential~\cite{fenepotential} for separations $\bar{r}_{ij}\geq 1$:  
\begin{equation}
\Phi_{cc}(\bar{r}_{ij}) = \Bigg\{ \begin{array}{cc} 
\epsilon (\bar{r}_{ij}^{-12}-2 \bar{r}_{ij}^{-6} +1) & \, \, \, \bar{r}_{ij} \leq 1 \\
-\epsilon \log{(1-q^{-2} (\bar{r}_{ij}-1)^2)} & \, \, \, \bar{r}_{ij} > 1 \\
\end{array}
\label{interaction1}
\end{equation}
where $\epsilon$ sets the energy scale and $q=0.1$.  This potential has a minimum of zero at $\bar{r}_{ij}=1$ and diverges at $\bar{r}_{ij}=1+q$ to prevent adjacent monomers from unbinding.
Green-green interactions are described by a Lennard-Jones (LJ) potential 
\begin{equation}
\Phi_{att}(\bar{r}_{ij}) = \epsilon E_c (\bar{r}_{ij}^{-12} - 2 \bar{r}_{ij}^{-6})
\label{interaction2}
\end{equation}
with energy depth $E_c<0$ at $\bar{r}_{ij}=1$, whereas green-white and white-white interactions obey a RLJ potential
\begin{equation}
\Phi_{rep}(\bar{r}_{ij}) = \Bigg\{ \begin{array}{cc} 
\epsilon (\bar{r}_{ij}^{-12}-2 \bar{r}_{ij}^{-6} +1) & \, \, \, \bar{r}_{ij} \leq 1 \\
0 & \, \, \, \bar{r}_{ij} > 1 \\
\end{array}
\label{interaction3}
\end{equation}
that provides a repulsive force when particles overlap and no force when they do not overlap.  

Thermal fluctuations are included using off-lattice Brownian dynamics simulations~\cite{browniandynamicssim}.  The vector position $\vec{x}_i$ of each monomer $i$ is determined at each time-step by the attractive and repulsive forces arising from the potentials in Eqs.~\ref{interaction1}-\ref{interaction3} and random forces arising from thermal fluctuations.  The equation of motion for monomer $i$ is
\begin{equation}
m_i \frac{d^2 \vec{x}_i}{d^2 t} = \vec{F}_i(t)-\eta \vec{v}_i-\frac{d}{d \vec{x}_i} \sum_{j \ne i} \big[(\Phi_{cc}(\bar{r}_{ij})+\Phi_{att}(\bar{r}_{ij}) + \Phi_{rep}(\bar{r}_{ij})\big],   
\label{newteqn}
\end{equation}
where $\vec{F}_i(t)$ is a Gaussian random force and $-\eta \vec{v}_i$ a damping force, with $\vec{v}_i$ denoting the velocity of monomer $i$ and $\eta$ the solvent viscosity.  The Gaussian random force has zero mean and a standard deviation proportional to $T/\eta$.  We solve Eq.~\ref{newteqn} using standard numerical integration techniques~\cite{browniandynamicssim} in the limit that monomer mass $m_i=0$.

Folding simulations are conducted by starting with $E_c=0$ and decreasing $E_c$ linearly in time with rate $r$ at constant $T=1$.  In the supporting information we include two movies from our simulations.  These show the behavior of a two dimensional polymer chain at $r \eta \sigma^2/T = 10^{-7}$ where folding occurs reliably (``slowrate.mov'') and at $r \eta \sigma^2/T = 10^{-5}$ where a misfold occurs (``fastrate.mov'').  
 
\subsection{Calculating energy landscapes and free energy}
The energy landscape in Fig.~\ref{energylandscape} is created by running $20$ separate folding simulations at each of five rates $r \eta \sigma^2 /T = \,10^{-8},\, 10^{-7}, \,10^{-6}, \,10^{-5},\, \mathrm{and} \,10^{-4}$.  Each simulation explores the range $0<c<0.4$ and the energy landscape is obtained by constructing a histogram over all observed states.  We believe that the landscape is sufficiently sampled since we observe that there is very little difference at small $D$ and $R_g$ between the energy landscape pictured in Fig.~\ref{energylandscape} and ones measured using only data from the smallest $r$.

The free energies in Fig.~\ref{figfreeenergya} are measured by slowly ramping to the desired $c$-value with $r \eta \sigma^2/T=5 \times 10^{-9}$, and then calculating a histogram of the probability $P(E,D)$ to have energy $E$ and end-to-end distance $D$ over $10^8$ time-steps for each $c$-value reported.  The free energy $F(E,D)$ is determined (within an additive constant) from the probability via the relation $F(E,D) = -T \log P(E,D)$.  

\subsection{Calculating $r^f$ and $r^s$}
The limiting rates can be determined using equations similar to those in Eqs.~\ref{tempfall} and \ref{tempstuck},   
\begin{eqnarray}
r^f = (c_3-c_2) r^* \exp(\frac{-\Delta F}{T}), 
\label{equationrf}
\\
r^s = \int_{c_3}^{\infty} r^* \, \exp(\frac{-\Delta F'(c)}{T}) \, dc.
\label{equationrs}
\end{eqnarray}
These equations are derived for the simulation protocol where $|E_c| = c T$ increases linearly in time to induce folding, with $T$ constant.  The maximum waiting time is taken to infinity.  

We first calculate $r^f$.  The data in Fig.~\ref{figfreeenergya} gives $c_2=0.0085$ and $c_3=0.01$.  The free energy barrier $\Delta F/T$ is determined by preparing the protein in topology $\mathbf{t}^5$ at $c=c_2$ and measuring the amount of time $t_f$ required to transition to topology $\mathbf{t}^0$, averaged over $100$ trials.  The free energy barrier is related to the transition time by $t_f = \exp(\Delta F/T)/r^*$.  We measure $t_f T/\eta \sigma^2= 8400$, where $\eta \sigma^2/T$ is the fundamental unit of time in the simulations.  Inserting these numbers into Eq.~\ref{equationrf} yields $r^f \eta \sigma^2/T = 1.8 \times 10^{-7}$.

\begin{figure}[!h]
\begin{center}
\mbox{
\psfrag{yl}{\Huge{$\log(t_s(c)  \,\,T /\eta \sigma^2)$}}
\psfrag{xl}{\Huge{$c$}}
\scalebox{0.24}{\includegraphics{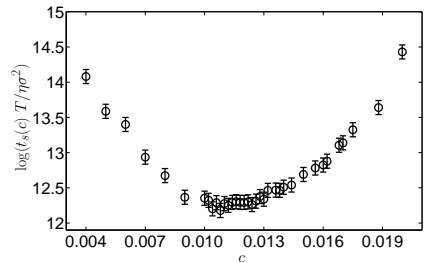}}
}
\end{center}
\caption{\label{figfreeenergyb}
Average time to transition from $\mathbf{t}^3$ to $\mathbf{t}^5$ as a function of $c$.  
}
\end{figure}

The rate $r^s$ is determined by preparing the protein in topology $\mathbf{t}^3$ and measuring the average time $t_s(c)$ required to transition to the native topology $\mathbf{t}^5$.  We average $t_s(c)$ over $100$ trials for each $c$-value and it is plotted in Fig.~\ref{figfreeenergyb}.  Since $t_s(c) = \exp(\Delta F'(c)/T)/r^*$ we calculate $r^s \eta \sigma^2/T = 3.0 \times 10^{-8}$ by direct integration of $t_s(c)^{-1}$, according to Eq.~\ref{equationrs}.  Contributions to the numerical value of $r^s$ from $c>0.02$ are negligible.

\begin{acknowledgments}
Financial support from NSF grant numbers CBET-0348175 (GL,JB), 
DMR-0448838 (GL,CSO), and Yale's Institute for Nanoscience and Quantum Engineering (GL) is gratefully acknowledged.
 We also thank Yale's High Performance Computing Center for computing time.
\end{acknowledgments}

\clearpage
\appendix*
\section{Supporting Information: Simulation results in three dimensions}
In the manuscript, simulation results were presented for a two dimensional model protein.  Here we include results for three dimensions.  These results exhibit similar behavior and support the theoretical predictions.

\begin{figure*}[p]
\begin{center}
\scalebox{0.80}{\includegraphics{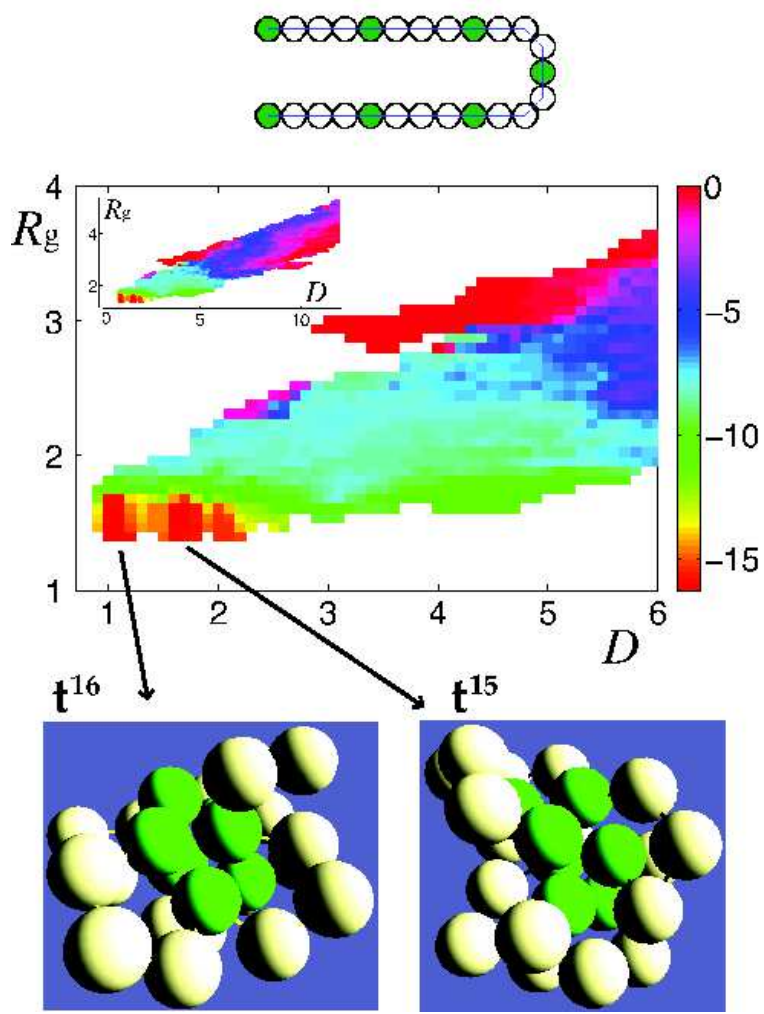}}
\end{center}
\caption{\label{3denergylandscape} Energy landscape and relevant topologies for a three dimensional model protein, pictured in an extended state with no bonds at the top of the figure.  The inset is the full energy landscape, and the main figure contains a magnified view of the compact states.  The colorbar gives the total energy of the system normalized by the magnitude of the attraction strength $|E_c|$.  There are two distinct energy minima separated by barriers and the topologies of each minima are pictured and labeled.  White regions correspond to protein conformations that are never sampled in the simulations.}
\end{figure*}

We perform off-lattice Brownian dynamics simulations in three dimensions to simulate the folding process.  We study the model protein pictured in Fig.~\ref{3denergylandscape} that consists of $25$ monomers, seven of which are attractive.  In Fig.~\ref{3denergylandscape} we plot the protein energy landscape as a function of the radius of gyration $R_g$ and end-to-end distance $D$, each normalized by the monomer diameter $\sigma$.  There are two minima at small $R_g$ and $D$, corresponding to the topologies $\mathbf{t}^{15}$ and $\mathbf{t}^{16}$ pictured in the figure.

\begin{figure*}[p]
\begin{center}
\begin{tabular}{c}
{\bf (a)} \\
\mbox{
\psfrag{t15}{\Huge{$\mathbf{t}^{15}$}}
\psfrag{t16}{\Huge{$\mathbf{t}^{16}$}}
\psfrag{yl}{\Huge{$E/|E_c|$}}
\psfrag{xl}{\Huge{$c$}}
\scalebox{0.40}{\includegraphics{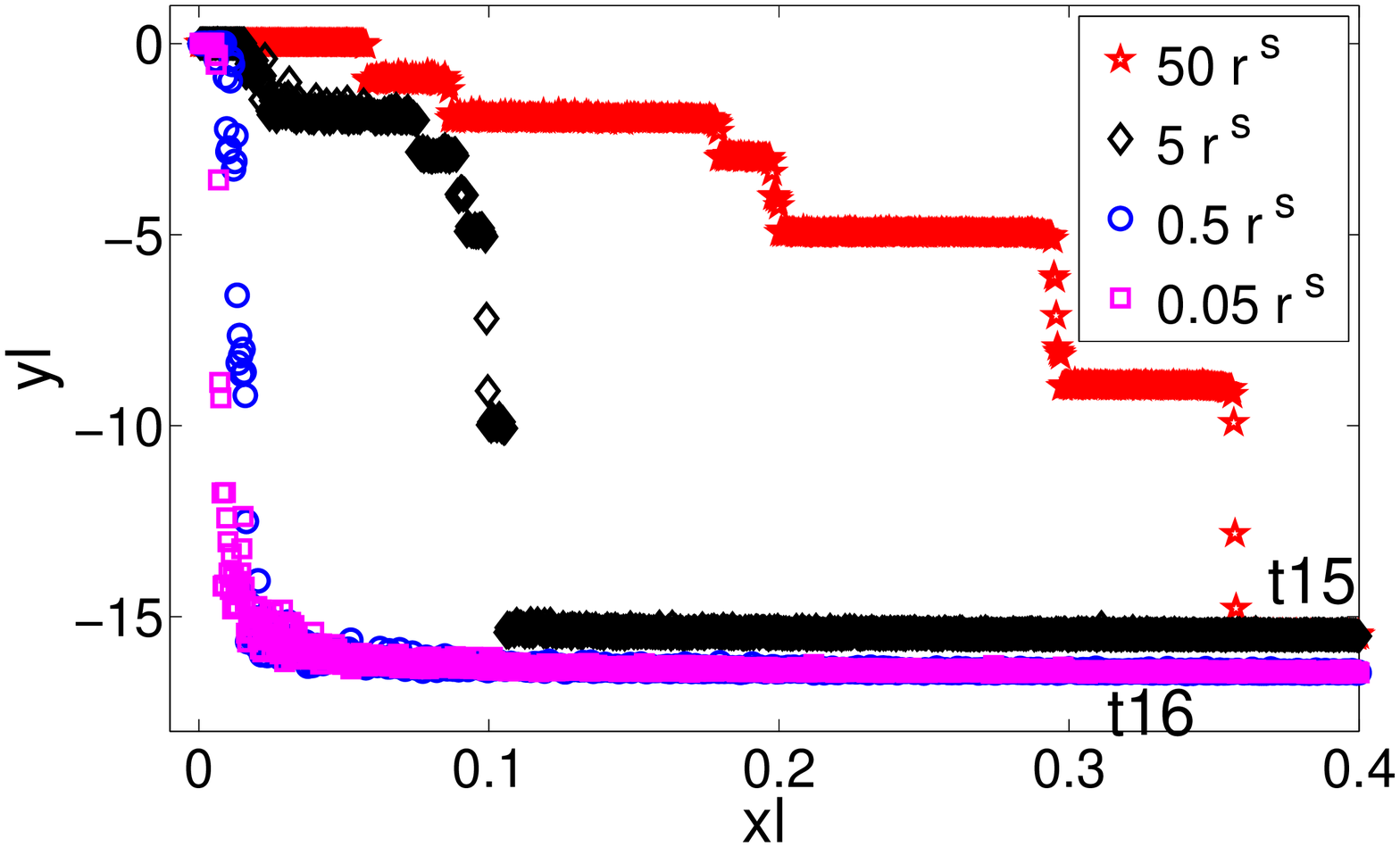}}}
\\
{\bf (b)} \\
\mbox{
\psfrag{xl}{\Huge{$\log_{10}{(r\eta \sigma^3/T)}$}}
\psfrag{pp}{\Huge{$P_c$}}
\scalebox{0.40}{\includegraphics{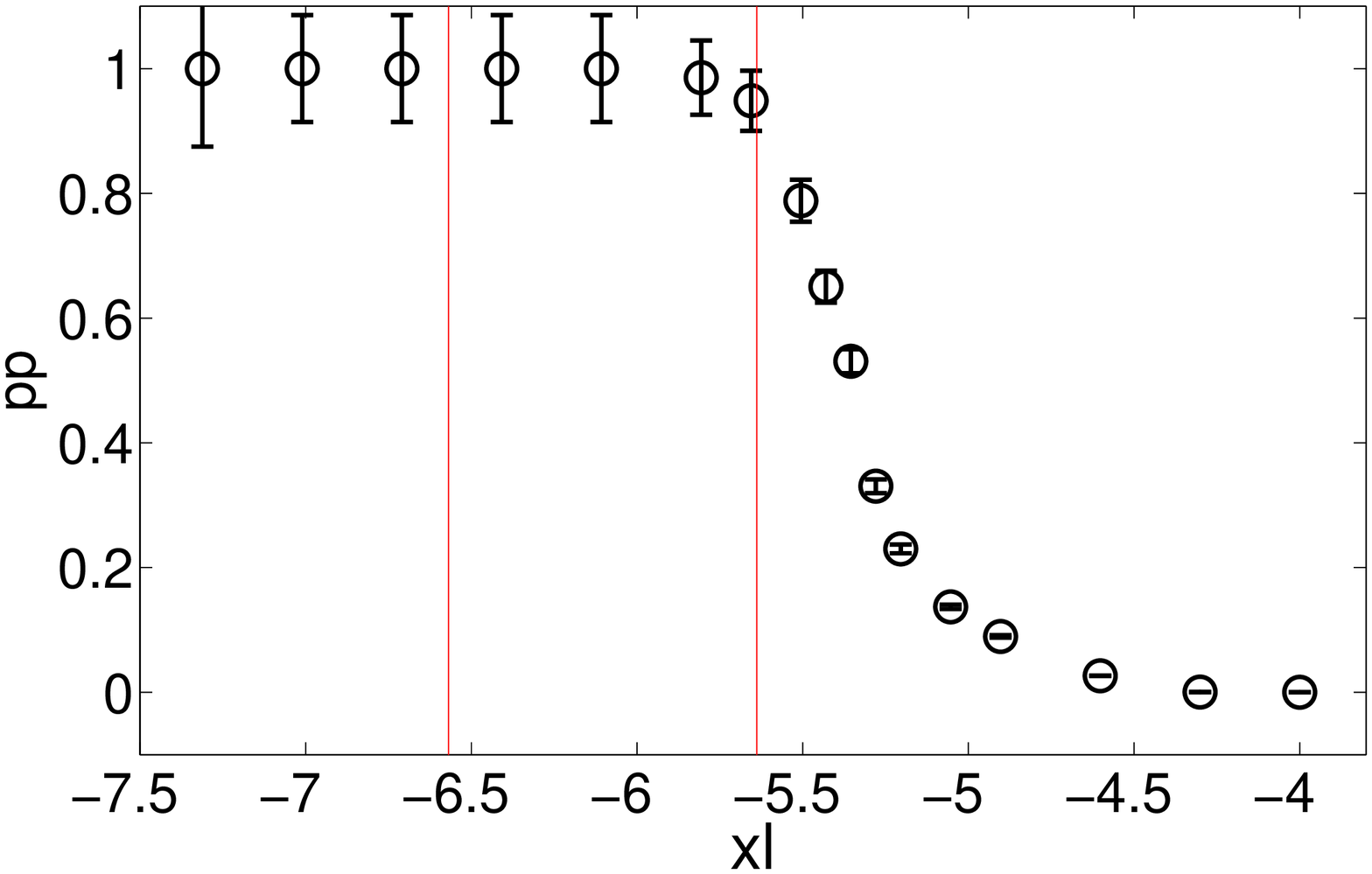}}
}
\end{tabular}
\end{center}
\caption{\label{3dfigfolding}  {\bf (a)} Folding trajectories in simulations with identical initial conditions at four different rates.  The normalized energy $E/|E_c|$ is plotted as a function of c and the final state is labeled by its topology.  Slow rates find the native state $\mathbf{t}^{16}$ reliably whereas fast rates give rise to unreliable folding.  {\bf (b)} The probability $P_c$ of folding to the native state $\mathbf{t}^{16}$ as a function of rate $r$.  Error bars are from sampling statistics.  For $r \eta \sigma^3 / T \lesssim 2 \times 10^{-6}$ the system folds reliably.  Vertical lines indicate the values of $r^f$ and $r^s$ calculated in the text.
}
\end{figure*}

As in the two dimensional case, non-funneled energy landscapes promote misfolding if the rate that the attractive strength $|E_c|$ is increased to induce folding is sufficiently large.  In Fig.~\ref{3dfigfolding}(a) we plot the energy as a function of $c \equiv |E_c|/T$.  For small rates the simulated protein folds to the global energy minimum $\mathbf{t}^{16}$.  For larger rates the system misfolds to the local minimum $\mathbf{t}^{15}$.  In Fig.~\ref{3dfigfolding}(b) we plot the probability to fold to the native state $\mathbf{t}^{16}$ as a function of rate.  The protein folds reliably below a normalized rate of $\sim 2 \times 10^{-6}$.

\begin{figure*}[p]
\begin{center}
\mbox{
\scalebox{0.55}{\includegraphics{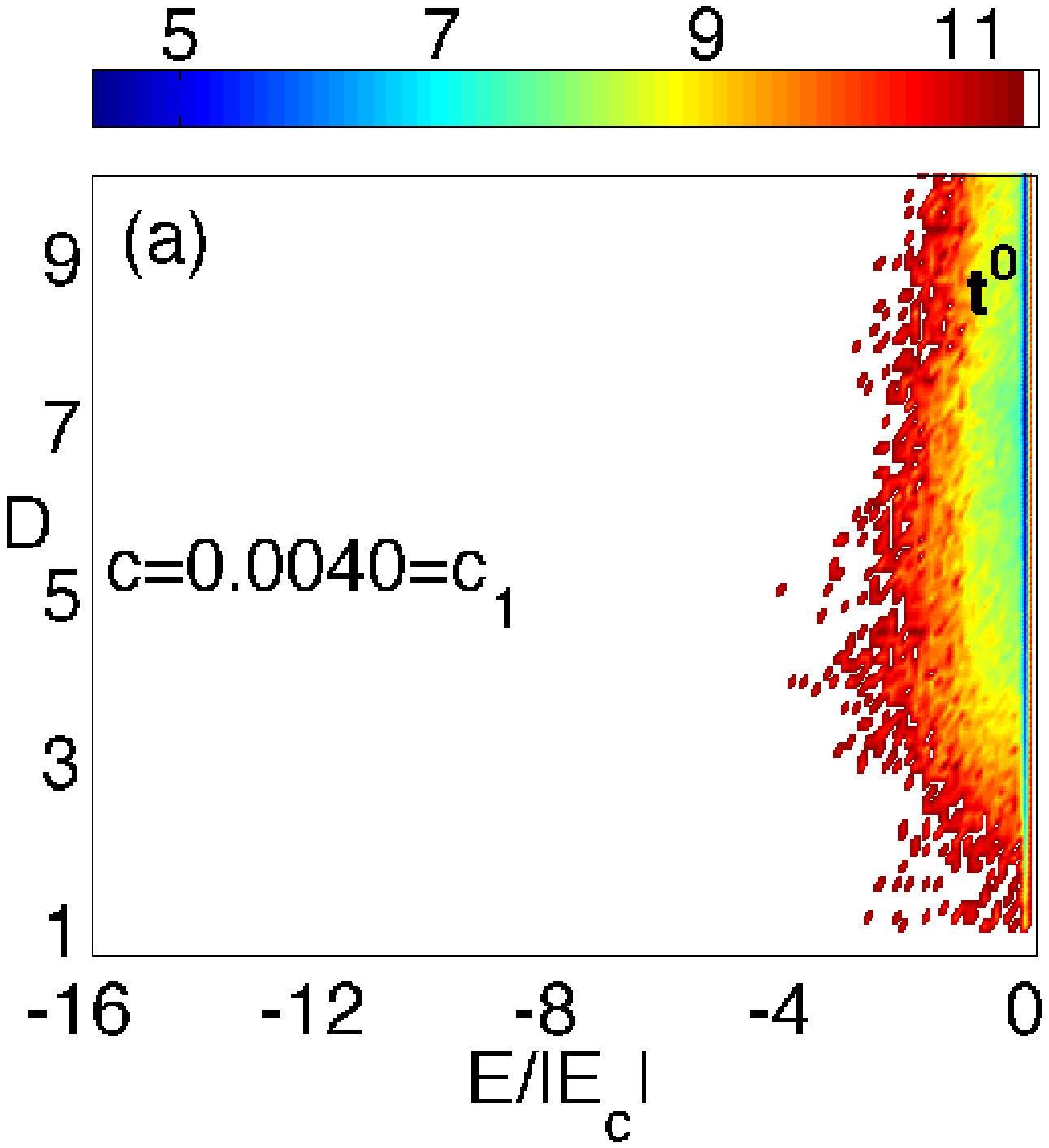}}
}
\mbox{
\scalebox{0.55}{\includegraphics{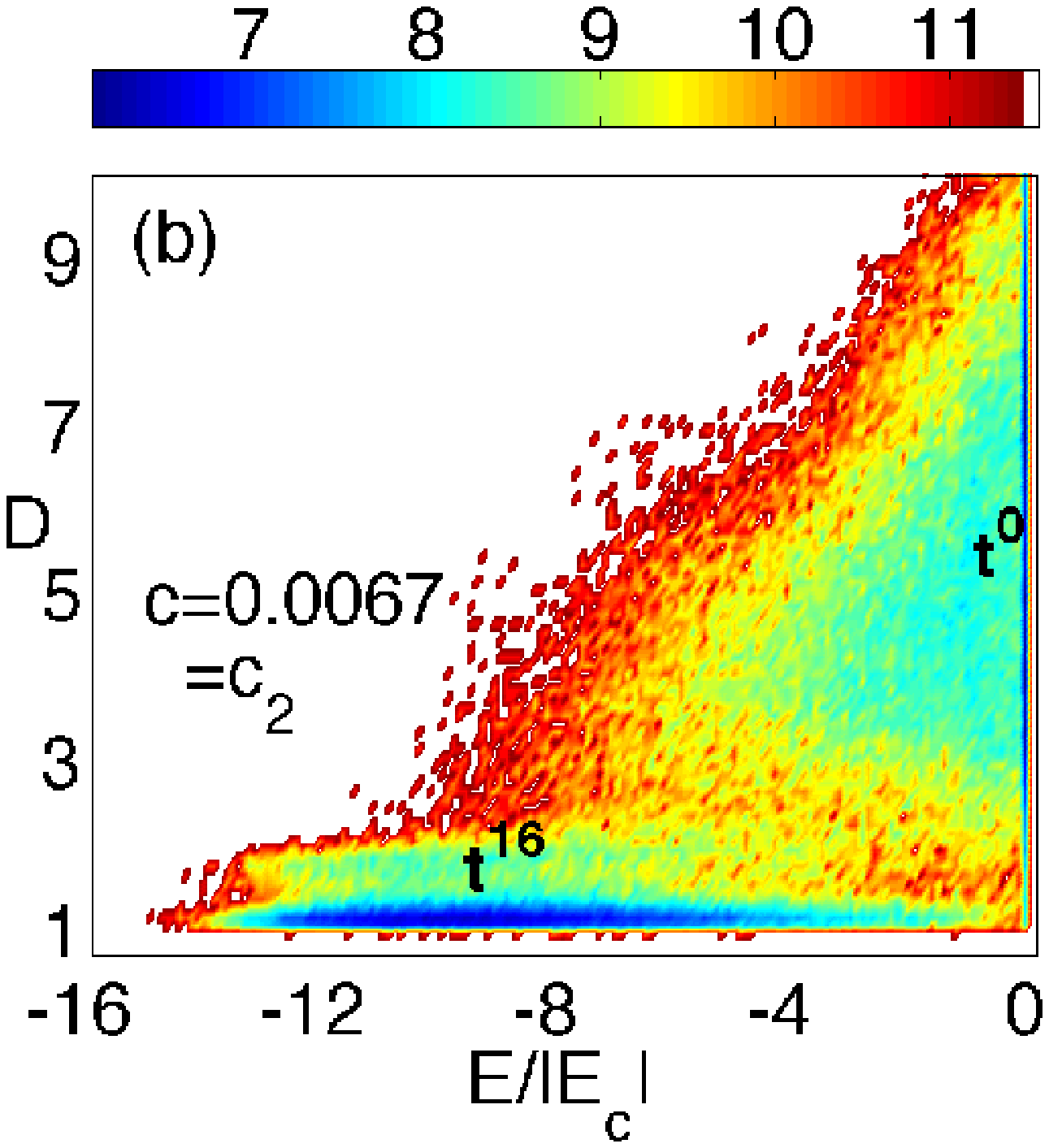}}
}
\mbox{
\scalebox{0.55}{\includegraphics{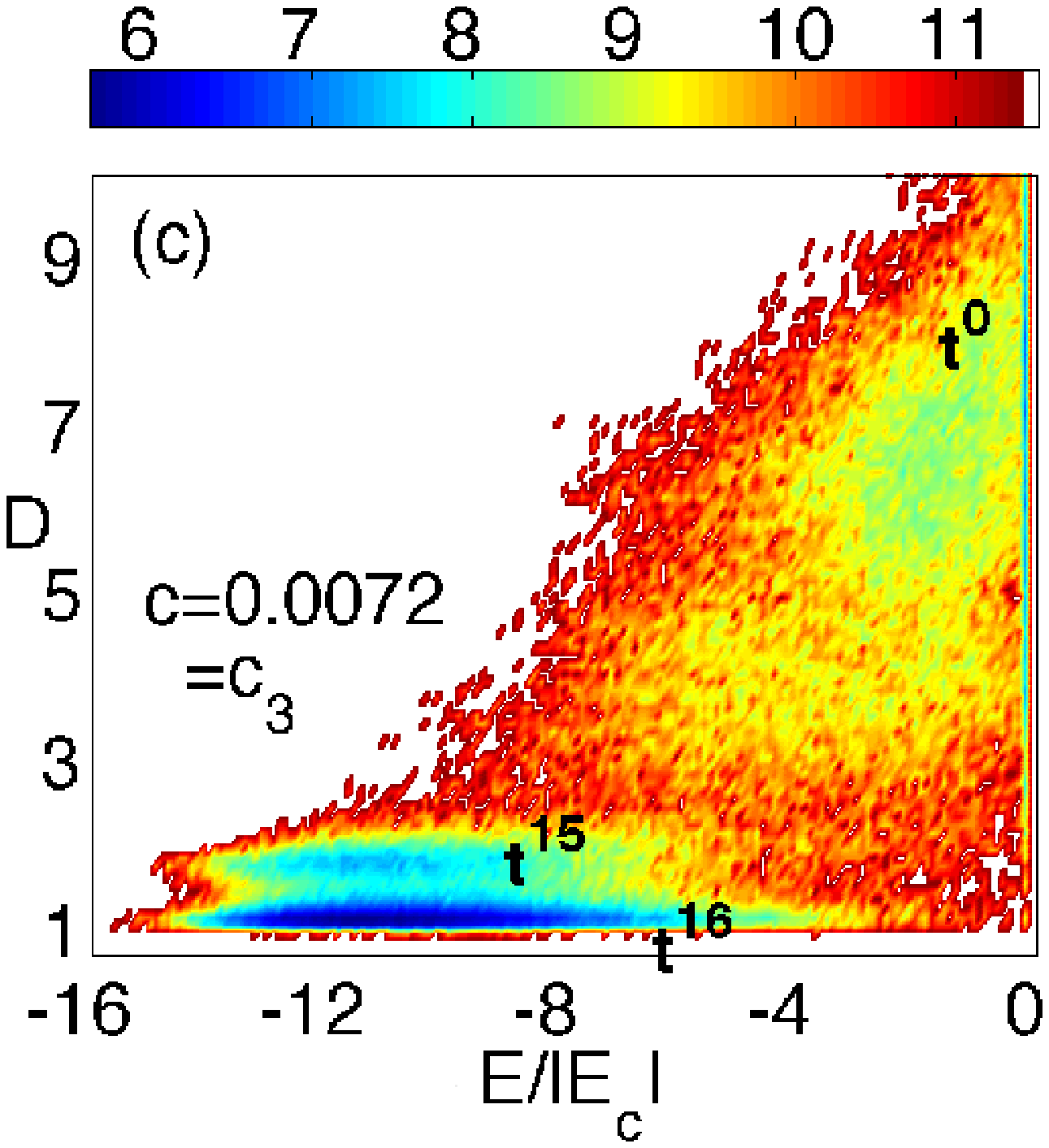}}
}
\mbox{
\scalebox{0.55}{\includegraphics{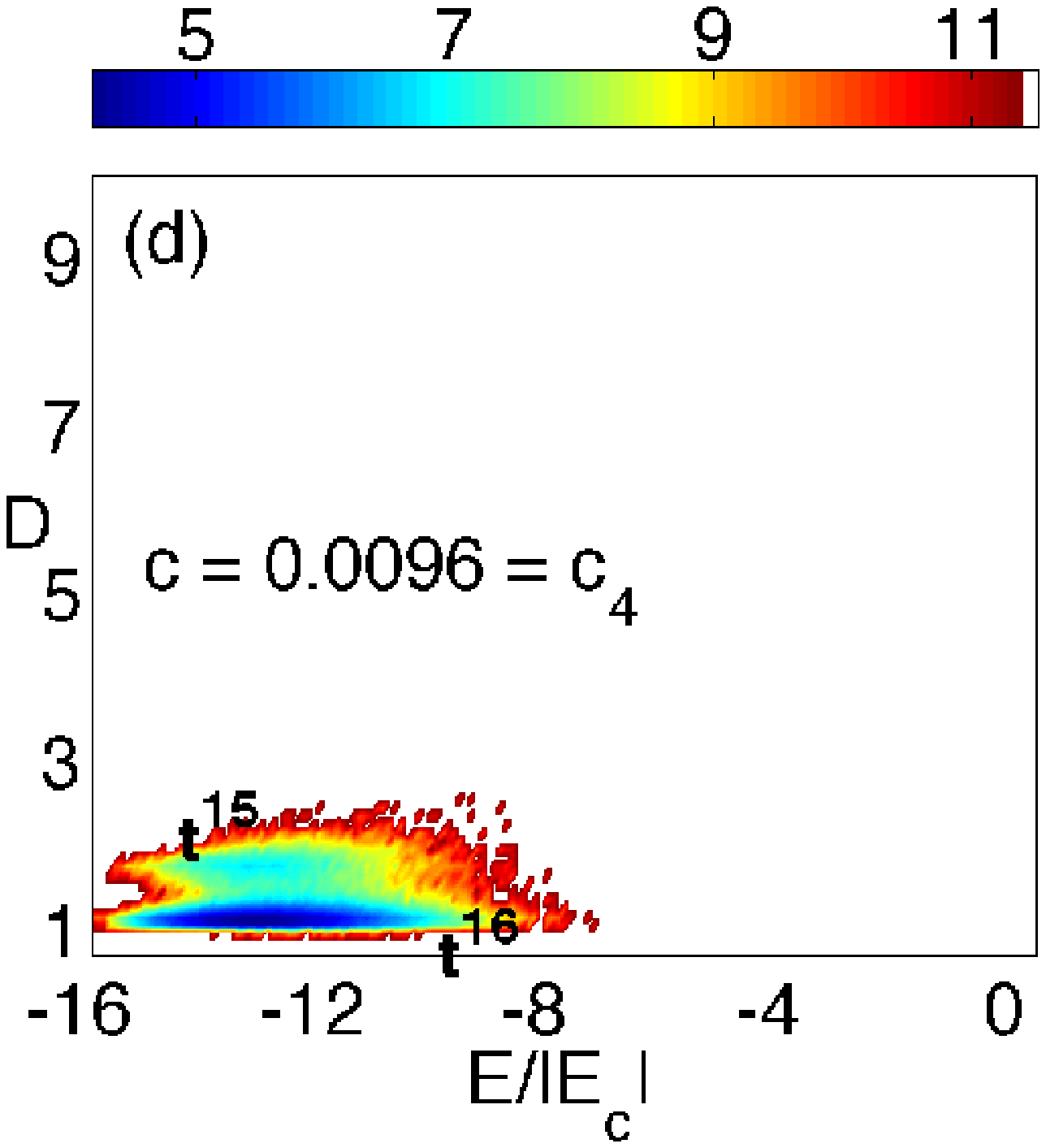}}
}
\end{center}
\caption{\label{figfreeenergy}  Contour plots of the free energy $F/T$ normalized by the temperature, as a function of the normalized energy $E/|E_c|$ (horizontal axis) and end-to-end distance $D$ (vertical axis) for four values of $c$.  White regions correspond to protein conformations that are never sampled in the simulations.
}
\end{figure*}

The limiting rate below which folding is reliable can be predicted by measurements of free energy.  In Fig.~\ref{figfreeenergy} we plot the free energy as a function of end-to-end distance $D$ and normalized energy $E/|E_c|$ for many different values of $c$.  In Fig.~\ref{figfreeenergy}(a) the random coil state $\mathbf{t}^0$ is the only minimum in the free energy.  For $c=0.0067$, Fig.~\ref{figfreeenergy}(b) demonstrates that $\mathbf{t}^{16}$ and $\mathbf{t}^{0}$ have equal free energies.  In Fig.~\ref{figfreeenergy}(c) the random coil $\mathbf{t}^0$, native state $\mathbf{t}^{16}$, and metastable state $\mathbf{t}^{15}$ basins of attraction are present.  At this value of $c=0.0072$, topology $\mathbf{t}^{15}$ has a free energy equal to that of $\mathbf{t}^0$.  For larger $c$ Fig.~\ref{figfreeenergy}(d) demonstrates that the protein has an increasing probability to populate the basin of attraction for $\mathbf{t}^{16}$, although the basin of attraction for $\mathbf{t}^{15}$ is still visible.  From this series of free energy plots, it is apparent that the simulated protein possesses a single equilibrium transition at $c=c_2$ from $\mathbf{t}^0$ to $\mathbf{t}^{16}$, and misfolds to $\mathbf{t}^{15}$ are possible for $c>c_3$.

The rate $r^f$ is calculated using the values $c_2=0.0067$ and $c_3=0.0072$, along with the transition time $t_f$ from $\mathbf{t}^{16}$ to $\mathbf{t}^0$ at $c=0.0067$.  We measure $t_f T/\eta \sigma^3 = 1850$, averaged over one hundred trials.  Given these values we calculate $r^f \eta \sigma^3/T=2.7 \times 10^{-7}$.

\begin{figure*}[p]
\begin{center}
\mbox{
\psfrag{yl}{\Huge{$\log(t_s(c) T/\eta \sigma^3$}}
\psfrag{xl}{\Huge{$c$}}
\scalebox{0.40}{\includegraphics{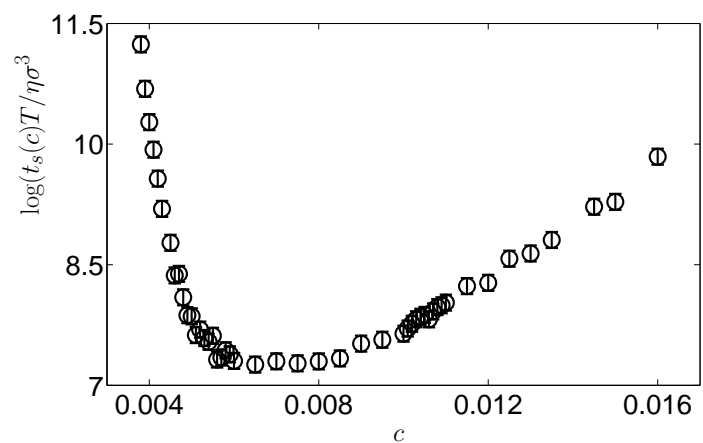}}
}
\end{center}
\caption{\label{freeenergybarrier} Average time to transition between $\mathbf{t}^{15}$ and $\mathbf{t}^{16}$ as a function of $c$.
}
\end{figure*}

The rate $r^s$ is calculated by measuring the transition time $t_s(c)$ between topologies $\mathbf{t}^{16}$ and $\mathbf{t}^{15}$, which is shown in Fig.~\ref{freeenergybarrier}.  Directly integrating this data for $c>c_3$ yields $r^s \eta \sigma^3/T = 2.3 \times 10^{-6}$.

Given the values of $r^f$ and $r^s$ we expect the protein to fold reliably for $r\eta \sigma^3/T < 2.3 \times 10^{-6}$, which is consistent with the data in Fig.~\ref{3dfigfolding}(b).
In contrast to the two dimensional simulations, we find $r^f < r^s$ and thus this particular protein can only fold in equilibrium.  Generally we believe that the ordering of $r^f$ and $r^s$ can depend on the length, sequence and energy scales of the protein.

\end{document}